\documentclass[osajnl,showpacs,superscriptaddress,11pt]{revtex4-1} 
\usepackage{amsmath,amssymb,graphicx,graphics,subfigure}
\usepackage{epstopdf}
\usepackage{verbatim} 

\newcommand{\zl}[2]{$#1\:\text{#2}$}

\newcommand{\be}{\begin{equation}}
\newcommand{\ee}{\end{equation}}
\newcommand{\bp}{\begin{pmatrix}}
\newcommand{\ep}{\end{pmatrix}}
\newcommand{\bea}{\begin{equation*}}
\newcommand{\eea}{\end{equation*}}

\begin{document}

\title{Simple method for locking birefringent resonators}



\author{Adam Libson}\email{Corresponding author: alibson@ligo.mit.edu}
\author{Nicolas Brown}
\author{Aaron Buikema}
\author{Camilo Cela L\'{o}pez}
\author{Tamara Dordevic}
\author{Matthew Heising}
\author{Matthew Evans}
\affiliation{LIGO Laboratory, Massachusetts Institute of Technology, Cambridge, Massachusetts 02139, USA}
\begin{abstract}
We report on a simple method of locking a laser to a birefringent cavity using polarization spectroscopy.  The birefringence of the resonator permits the simple extraction of an error signal by using one polarization state as a phase reference for another state.  No modulation of the light or the resonator is required, reducing the complexity of the laser locking setup.  This method of producing an error signal can be used on most birefringent optical resonators, even if the details of birefringence and eigenpolarizations are not known.  This technique is particularly well suited for fiber ring resonators due to the inherent birefringence of the fiber and the unknown nature of that birefringence. We present an experimental demonstration of this technique using a fiber ring.
\end{abstract}


\maketitle 


\section{Introduction}
Using optical resonators to measure and control the frequency of a laser is a well-established technique. The use of locking methods in which multiple modes of a probe light field are compared are especially effective. Examples include the Pound-Drever-Hall (PDH) locking scheme \cite{Pound1946,Drever1983}, in which different frequency components are interfered, and schemes in which different spatial modes are compared \cite{Shaddock1999,Miller2014}. In general, any two orthogonal modes can be used to make a phase-sensitive error signal \cite{Harvey2003}. In this paper, we generalize the locking technique of H\"ansch and Couillaud (HC) \cite{Hansch1980}, which utilizes different polarization modes to produce an error signal. 
The HC method uses a polarizer in the resonator to produce a polarization-dependent resonance condition. Variations of this technique have been demonstrated in free space cavities by introducing a birefringent crystal into the cavity \cite{Boon1997}, using a non-planar ring cavity \cite{Honda2009}, using a triangular cavity \cite{Moriwaki2009}, or using the birefringence of dielectric mirrors \cite{Asenbaum2011}.  We generalize these locking techniques and describe and demonstrate a polarization spectroscopy locking method that introduces no losses, is simple to implement, and requires no knowledge of the cavity birefringence. These properties make this method particularly attractive for systems with inherent or unknown birefringence, such as whispering-gallery-mode (WGM) resonators \cite{Alnis2011,Schliesser2008}, cavities formed with crystal-coated mirrors \cite{Cole2013,Kessler2012}, and fiber-based applications.

The resonance properties \cite{Zhang1988} and the polarization effects of fiber rings have been investigated previously \cite{Lamouroux1982, Iwatsuki1986, Wang2011}. Small stresses and imperfections in the core of the fiber produce polarization-dependent phase shifts that are rarely known \emph{a priori}. This inherent unknown birefringence make this locking technique well suited to use with fiber ring resonators. In addition to the ease of alignment and mode-matching that comes with using single-mode optical fibers, fiber rings allow for long cavities and narrow linewidths in compact packages with minimal back reflection to the laser source. Traditional  methods such as PDH can be used to lock to fiber rings \cite{Merrer2008}, but there are techniques unique to fiber optics, such as using Rayleigh scattering to provide optical feedback \cite{Paul1993}. In contrast, the locking scheme we present requires no modulation or demodulation, which allows for simple electronics and a high-bandwidth error signal. Further, it avoids the challenges of optical feedback. We demonstrate this technique experimentally with a fiber ring. This method can be used to produce compact pre-stabilized lasers for use in fiber-based systems that require stable coherent sources, such as telecommunication systems \cite{Brinkman2002}, lidar \cite{Ostermeyer2005}, fiber gyroscopes \cite{Lefevre1982}, and other fiber-based sensors.

\section{Polarization in Optical Resonators} \label{sec:resonators}

Most optical resonators will have some effect on the polarization of circulating light. A birefringent cavity will impart a different phase delay to each polarization and may alter the polarization of the input light. Additionally, cavity losses may be polarization dependent, which can create additional interesting polarization effects.

These polarization effects can be more easily understood by representing the electric field of fully polarized light as a two-component complex vector known as a Jones vector. Conventionally, this vector is written using linear horizontal and vertical polarizations as the basis, so a polarized plane light wave at time $t$ a distance $z$ along the propagation axis will have the form
\begin{equation}
\vec{E} = \begin{pmatrix} E_\text{H} \\ E_\text{V}\end{pmatrix} =  \begin{pmatrix} A_\text{H} \\ A_\text{V}\text{e}^{i\phi}\end{pmatrix} \text{e}^{i(\omega t-kz)},
\end{equation}
where $A_\text{H}$ and $A_\text{V}$ are the amplitudes of each polarization component, $\phi$ is the phase difference between each component, and $\omega$ and $k$ are the angular frequency and wavenumber of the light, respectively. The last exponential is usually dropped since global phases have no effect on the polarization. In this framework, cavity losses, polarization transformations, and phase changes can be described by a $2 \times 2$ Jones matrix that acts on the polarization vector. For example, in this basis the Jones matrix for a half-wave plate with fast axis at angle $\theta$ to the horizontal is 
\begin{equation}
\begin{pmatrix}
\cos 2 \theta & \sin 2 \theta \\ 
\sin 2 \theta & - \cos 2 \theta
\end{pmatrix}.
\end{equation}

\begin{figure}
\centerline{\includegraphics[width=.8\columnwidth]{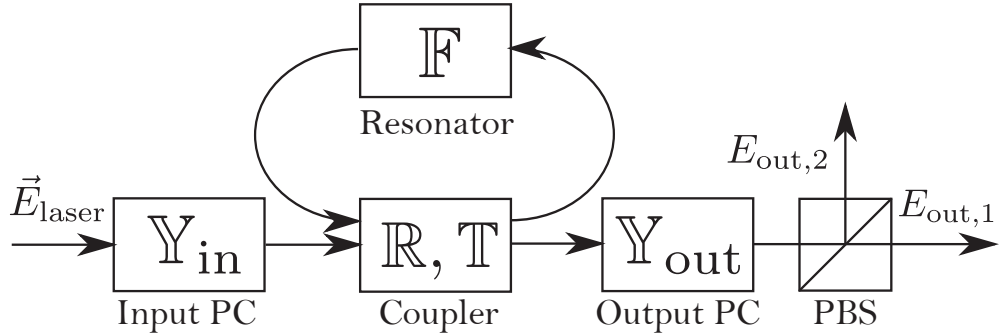}}
\caption{Block diagram of an optical experiment compatible with our locking method. Polarization controllers (PCs) on the input and output of the resonator are given by the Jones matrices $\mathbb{Y}_{\text{in}}$ and $\mathbb{Y}_{\text{out}}$. Matrices $\mathbb{R}$ and $\mathbb{T}$  represent reflection and transmission of the coupler, respectively, and $\mathbb{F}$ represents propagation through the resonator. A polarizing beam splitter (PBS) follows at the end.
\label{fig:block_diagram}}
\end{figure}

In what follows, we will consider the polarization effects of a fiber ring resonator with a single input coupler, but this method is completely generalizable to any birefringent cavity. The amplitude transmittance and reflectance matrices of the input coupler are labelled $\mathbb{T}$ and $\mathbb{R}$, respectively. Let $\mathbb{F}$ be the matrix that represents the polarization transformation as the light propagates through the fiber ring (see figure \ref{fig:block_diagram}). Included in this matrix is information about the overall (frequency-dependent) phase picked up in a single trip through the ring and any losses. The total cavity amplitude reflectivity matrix $\mathbb{R}_\text{cav}$, defined by $\vec{E}_\text{refl}=\mathbb{R}_\text{cav} \vec{E}_\text{in}$, is given by
\begin{equation}\label{eq:Rcav}
\mathbb{R}_\text{cav} = \mathbb{R} - \mathbb{T}  \mathbb{F}  \left(\mathbb{I} - \mathbb{R}  \mathbb{F} \right)^{-1}  \mathbb{T},
\end{equation}
where $\mathbb{I}$ is the identity matrix. Since $\mathbb{R}  \mathbb{F}$ represents one round trip of the resonator, the two eigenvectors of this matrix are the eigenpolarizations of the resonator. For positive real eigenvalues of $\mathbb{R}  \mathbb{F}$, light will resonate.

We assume that the eigenvectors of $\mathbb{RF}$ are also eigenvectors of $\mathbb{T}$ and $\mathbb{R}$, and therefore of $\mathbb{R}_\text{cav}$, for all frequencies $\omega$. This assumption holds for input couplers with polarization-independent properties. For birefringent input couplers, this assumption is valid when the eigenpolarizations of the coupler match those of the cavity.
For fibers, the polarization effects of the couplers are often much smaller than the effects due to birefringence in the fiber itself \cite{Burns1982}, so this is a reasonable assumption for fiber rings.

\begin{figure}
\centerline{\includegraphics[width=.7\columnwidth]{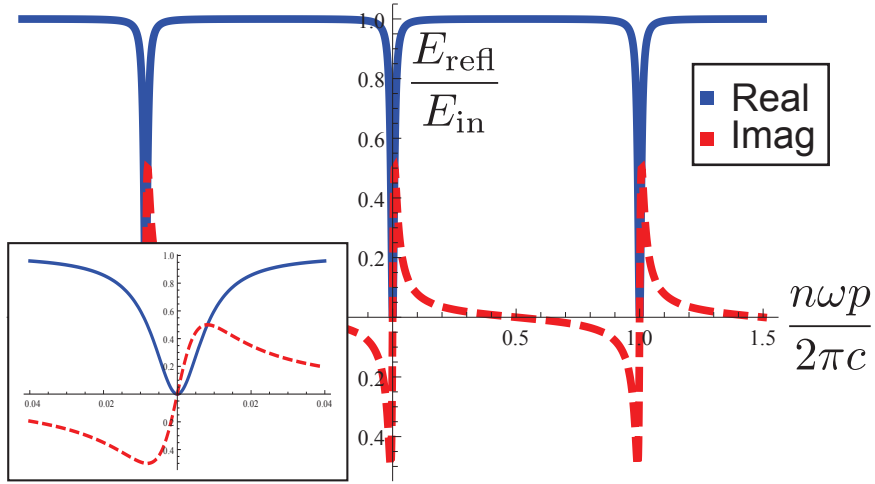}}
\caption{Cavity reflectance $r_\text{cav}$ as a function of frequency for a single input mode. Blue/Solid: Real, Red/Dashed: Imaginary. Input mirror reflectivity $r^2=1-t^2=0.95$, cavity losses $(1-\alpha^2)$ = 0.05. Inset: closeup of resonance peak. Note the sharp change in the imaginary component as the frequency moves through a cavity resonance.
\label{fig:dispersion}}
\end{figure}

In this case, the eigenvalues of $\mathbb{R}_\text{cav}$ are given by
\be
r^j_\text{cav}(\omega) = r_j - \frac{t_j^2 f_j(\omega)}{1-r_j f_j(\omega)},
\ee
where $t_j$, $r_j$, and $f_j$ are the eigenvalues of the $j$th eigenpolarization under $\mathbb{T}$, $\mathbb{R}$, and $\mathbb{F}$, respectively. In general, $f_j=\alpha_j e^{i k_j p}$, where  $k_{j}=2\pi n_{j}\nu/c$ is the wavenumber for each eigenpolarization,  $n_{j}$ is the effective index of refraction for each eigenpolarization, $\nu$ is the frequency of light, $p$ is the distance the light travels in one round trip of the cavity, and $\alpha_{j}$ is a complex term that accounts for losses in the cavity and additional phase shifts not due to propagation. Near resonance, the reflected light undergoes a very large phase shift. Figure~\ref{fig:dispersion} displays the real and imaginary parts of this reflection coefficient $r^j_\text{cav}(\omega)$ for a general optimally coupled cavity. Note the sharp change in the imaginary component near resonance due to the phase shift.

In general, the wavenumbers for different eigenpolarizations will not be equal, so there will be a difference in the phase accumulated by each eigenpolarization after one round trip, resulting in two longitudinal modes that are on resonance for different $\omega$. As one eigenpolarization moves through a resonance, it undergoes a large phase shift on reflection, while the other eigenpolarization does not. In this manner, the latter can be used as a phase reference for the former. This requires that the resonances to be shifted sufficiently far in frequency. Individual resonance peaks will be separated for round trip phase differences between eigenpolarizations $\Delta\theta$ that satisfy

\be
\Delta\theta\bmod{2\pi}>\frac{2\pi\Delta\nu}{\text{FSR}}
\ee
and
\be
-\Delta\theta\bmod{2\pi}>\frac{2\pi\Delta\nu}{\text{FSR}},
\ee
where $\Delta\nu$ is the full width at half maximum of the cavity resonance and \text{FSR} is the free spectral range. Thus, this method can be used even in cavities with small birefringence so long as the cavity has a large finesse ($\propto \frac{\text{FSR}}{\Delta\nu}$).

\section{The Error Signal}\label{sec:err_sig}

To use the birefringence of the resonator to produce an error signal, light travelling to the resonator must first be put into the correct input polarization, which can be done using a polarization controller (PC). There are many ways to implement such a device, but a common arrangement consists of a quarter-wave plate, a half-wave plate, and a second quarter-wave plate, each of which can be rotated independently. These polarization controllers can map an arbitrary input polarization state to any other polarization state \cite{Xiao-Guang2008}.  After reflecting from the resonator, light goes through another PC before propagating to a polarizing beam splitter (PBS) with a photodiode (PD) at each output (figure~\ref{fig:block_diagram}).  With the correct polarization control settings before and after the cavity, the difference in power at the two output ports of the PBS produces an error signal for the resonance condition of the cavity.  Expressed in the Jones matrix formalism the error signal $\Delta$ is proportional to 
\begin{equation}
\Delta \propto \lvert E_{\text{out},{2}} \rvert^{2} - \lvert E_{\text{out},{1}} \rvert^{2},
\end{equation}
where
\begin{align}
E_{\text{out},{1}} &= \begin{pmatrix} 1 \\ 0 \end{pmatrix} \cdot \mathbb{Y}_\text{out}  \mathbb{R}_\text{cav}  \mathbb{Y}_\text{in}  \vec{E}_\text{laser} \nonumber \\
E_{\text{out},{2}} &= \begin{pmatrix} 0 \\ 1 \end{pmatrix} \cdot \mathbb{Y}_\text{out}  \mathbb{R}_\text{cav}  \mathbb{Y}_\text{in}  \vec{E}_\text{laser}
\end{align}
are the respective projections of the final electric field amplitude onto the PBS polarization basis states. The proportionality constant will depend on the PD gain. In the above equations, $\mathbb{Y}_\text{in (out)}$ is the Jones matrix of the input (output) PC and $\mathbb{R}_{\text{cav}}$ is the cavity amplitude reflectivity matrix given by equation~\eqref{eq:Rcav}.  

We further assume the cavity eigenpolarizations are orthogonal. This will be true for cavities with polarization-independent losses, in which $\mathbb{R}$, $\mathbb{T}$, and $\mathbb{F}$ are all unitary matrices multiplied by a constant loss term \cite{Iwatsuki1986}. The eigenpolarizations will also be orthogonal for cavities in which one polarization mode is completely extinguished in the cavity, as in the original  H\"ansch-Couillaud method \cite{Hansch1980}. In practice, this approximation holds true for most single-mode fiber rings \cite{Iwatsuki1986}. 

Let $\vec{E}_a$ and $\vec{E}_b$ be the normalized eigenpolarizations of $\mathbb{R}\mathbb{F}$.  The input PCs are set so that the input polarization is an equal superposition of these eigenpolarizations:
\be\label{eq:in}
\vec{E}_\text{in} =\mathbb{Y}_\text{in}\vec{E}_\text{laser}= \frac{E_0}{\sqrt{2}}\left(\vec{E}_a + e^{i\gamma}\vec{E}_b\right),
\ee
where $E_0$ is the amplitude of the electric field and $\gamma$ is the phase difference between eigenpolarization components. The reflected light will be in the polarization state
\be
\begin{split}
\vec{E}_\text{refl} = \mathbb{R}_\text{cav}\vec{E}_\text{in}= \frac{E_0}{\sqrt{2}}\left(\mathbb{R}_\text{cav}\vec{E}_a + e^{i\gamma}\mathbb{R}_\text{cav}\vec{E}_b\right)\\
=\frac{E_0}{\sqrt{2}}\left(r^a_\text{cav}(\omega)\vec{E}_a + e^{i\gamma} r^b_\text{cav}(\omega)\vec{E}_b\right).
\end{split}
\ee
Because the polarization controllers can map an arbitrary input state to any given output polarization, there exists an arrangement of the output polarization controller that maps one eigenpolarization to an equal superposition of the PBS polarizations:
\be
\vec{E}_a\mapsto\mathbb{Y}_\text{out}\vec{E}_a =\frac{1}{\sqrt{2}} \bp 1\\e^{i\delta} \ep
\ee
in the PBS basis for some $\delta$. The polarization controller produces a lossless, and therefore unitary, transformation on the Jones vector, so the other eigenpolarization will be mapped to a final polarization orthogonal to this with some relative phase shift $\phi$:
\be
\vec{E}_b\mapsto\mathbb{Y}_\text{out}\vec{E}_b = \frac{e^{i\phi}}{\sqrt{2}}\bp -1\\e^{i\delta} \ep.
\ee
The final polarization state in the PBS basis is then
\begin{equation}
\vec{E}_\text{out} 
= \mathbb{Y}_\text{out}\vec{E}_\text{refl} 
= \frac{E_0}{2}\left(r^a_\text{cav}(\omega)\bp 1\\e^{i\delta} \ep + e^{i(\gamma+\phi)} r^b_\text{cav}(\omega)\bp -1\\e^{i\delta} \ep\right)
\end{equation}
and the sum and difference of the photodiode signals are proportional to
\be
\lvert E_{\text{out},{1}}\rvert^2+ \lvert  E_{\text{out},{2}}\rvert^2 =  \frac{E_0^2}{2}\left(\lvert r^a_\text{cav}(\omega)\rvert^2+\lvert r^b_\text{cav}(\omega)\rvert^2\right)\\
\ee
and
\begin{equation}  \label{eq:err_sig}
\lvert E_{\text{out},{2}}\rvert^2- \lvert  E_{\text{out},{1}}\rvert^2  =  E_0^2\: \text{Re}\left\{ (r^a_\text{cav}(\omega))^*  r^b_\text{cav}(\omega)e^{i(\gamma+\phi)} \right\},
\end{equation}
respectively. $\text{Re}$ denotes the real part and $^*$ represents complex conjugation.
Equation~\eqref{eq:err_sig} forms the error signal.

In the case of widely separated resonances, near the resonance of one eigenpolarization the imaginary part of the cavity reflectivity undergoes a sharp change, while the reflectivity for the other eigenpolarization changes very little. By adjusting the additional phase term in equation~\eqref{eq:err_sig}, the steep imaginary part of the individual reflection coefficients can be extracted. For widely separated resonances, this occurs for $\gamma + \phi\approx\pm\pi/2$. The output polarization controllers vary $\phi$ to produce an ideal error signal; therefore, it is not necessary to know or control $\gamma$. This means that the input polarization state can be chosen without regard to the relative phases of the eigenpolarizations, and that only the sum signal is needed to set the input PCs.

This is a general method for producing an error signal for a birefringent resonator with two longitudinal modes;  for a chosen input polarization satisfying \eqref{eq:in}, one can always find a physically realizable Jones matrix $\mathbb{Y}_\text{out}$ that produces an error signal.

Fundamentally, this method is limited only by shot noise, which scales as $P^{-1/2}$, where $P$ is the light incident on the cavity. In fiber systems, stimulated Brillouin scattering \cite{Ippen1972} puts a limit on the input power, which prevents the reduction of shot noise by turning up the optical power. However, in practice this method will be limited by acoustic/vibrational noise and thermorefractive effects in the fiber.

\section{Experimental Design and Data}\label{sec:exp}

\begin{figure}
\centerline{\includegraphics[width=.7\columnwidth]{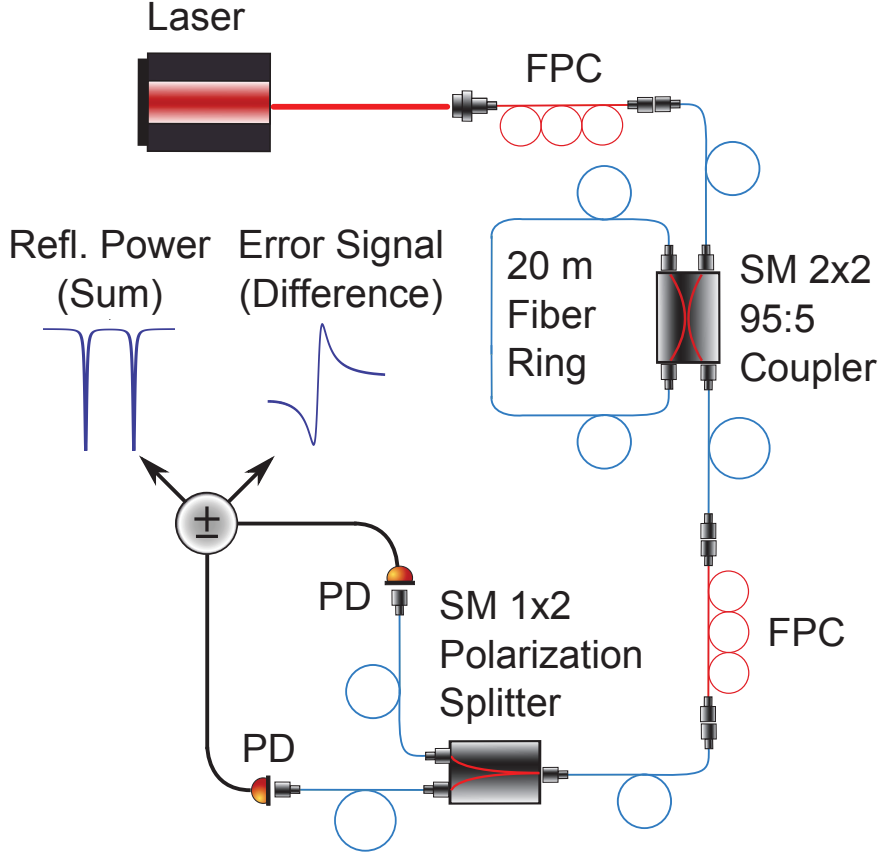}}
\caption{A schematic of the experimental setup used to produce the error signal.  The fiber polarization controllers (FPCs) approximate adjustable waveplates to tune the input and output polarizations to optimize the error signal.  The ${20}\text{-m}$ fiber ring has a finesse of $\approx60$. The sum signal is used to ensure equal input power in each eigenpolarization.
\label{fig:exp_setup}}
\end{figure}

The experimental setup needed to demonstrate this method is shown in figure~\ref{fig:exp_setup}.  We use a fiber ring resonator as the birefringent cavity. This trivializes cavity alignment and mode-matching.  Birefringence in optical fibers is often viewed as a trait that must be overcome \cite{Lamouroux1982}, but here we use it as a feature and exploit it to produce the error signal. We use a direct-coupling fiber ring setup \cite{Zhang1988} described below to produce the error signal.  

A non-planar ring oscillator (NPRO) Nd:YAG laser emitting at 1064 nm (Lightwave 125-1064-700) provides the incident light. The output frequency can be controlled by changing the temperature of the lasing medium crystal or by applying a strain to the crystal via a piezoelectric actuator. Approximately \zl{1}{mW} of linearly-polarized light is coupled into a single-mode (SM) optical fiber and encounters the first fiber polarization controller (FPC). Our controller consists of three rotating paddles with varying lengths of fiber coiled around each paddle. This approximates a quarter-wave plate, followed by a half-wave plate, and finally another quarter-wave plate.  Using this setup, we are able to map any input polarization to an arbitrary output polarization.

From there, the light enters an SM 95:5 $2\times2$ coupler spliced to a \zl{p=20}{m} length of single-mode fiber, forming a ring.  We expect approximately 4\% total losses in the splices and in the fiber itself.  This produces a nearly optimally coupled ring resonator with a finesse of approximately 60 for both eigenpolarizations and a free spectral range of \zl{c/np\approx 10}{MHz} \cite{Siegman1986} for a fiber core with index $n\approx1.6$. 

Following the coupler, the light again goes through an FPC, and then on to a fiber PBS.  Each output of the PBS goes to a photodiode, the outputs of which are subtracted using a low noise amplifier to produce the error signal. 

\begin{figure}
        \subfigure[]{%
            \label{fig:sum_data}
            \includegraphics[width=0.48\columnwidth]{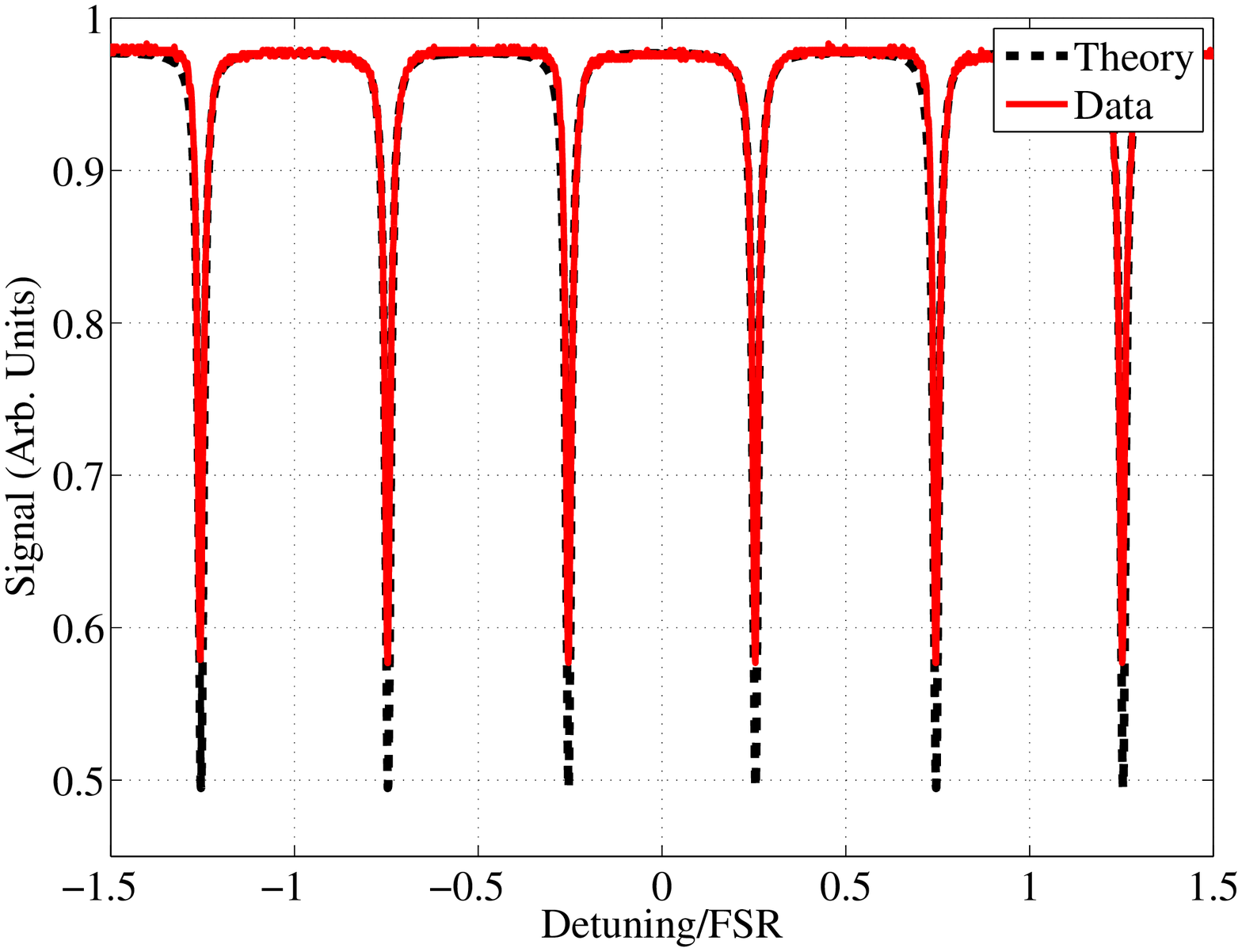}
        }%
	\subfigure[]{%
            \label{fig:error_data}
            \includegraphics[width=0.48\columnwidth]{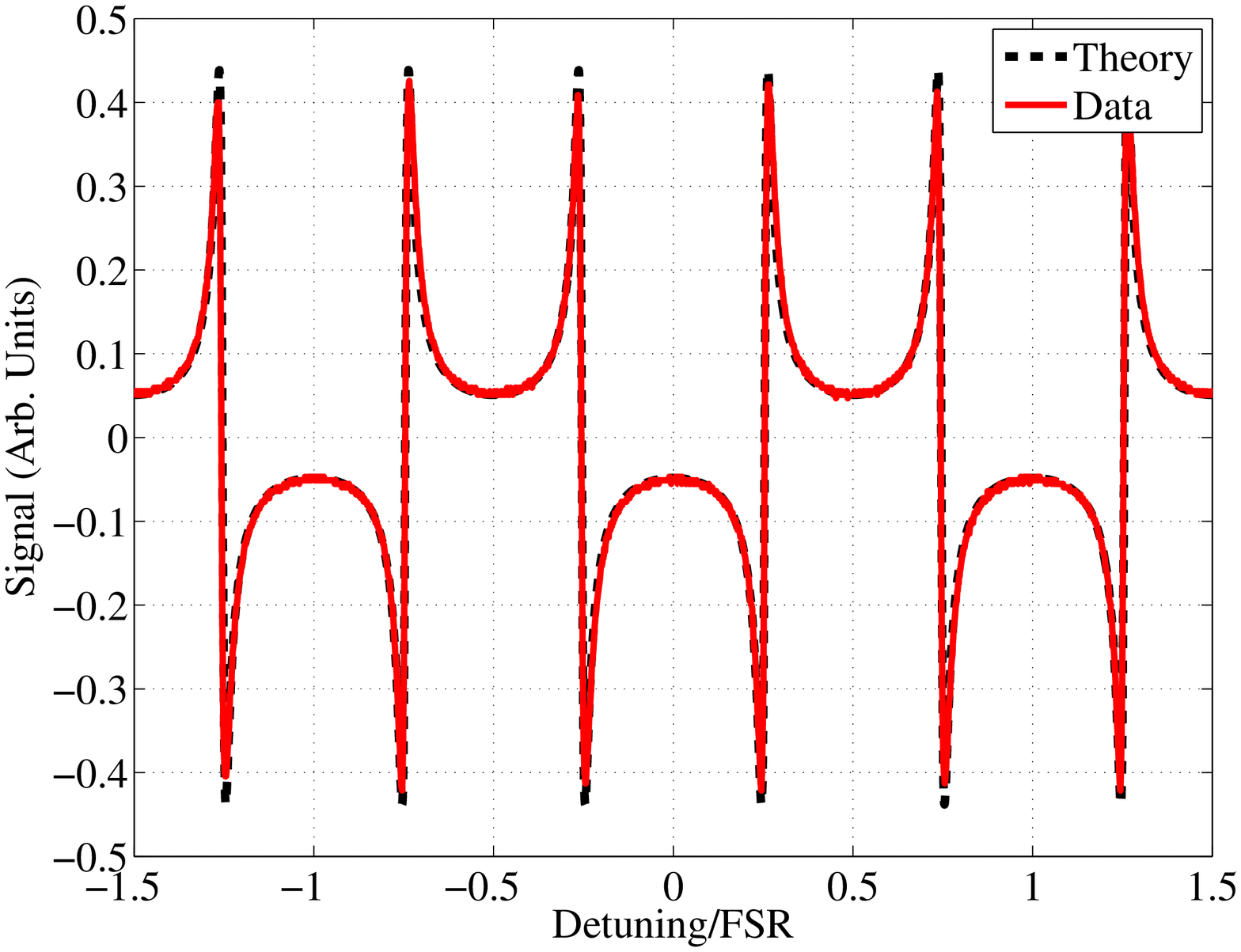}
        }%
    \caption{%
        Observed and modeled \subref{fig:sum_data} Sum and \subref{fig:error_data} difference of the two output ports of the PBS for the fiber ring resonator setup. The modeled resonator has a 95:5 coupler with 2\% loss and the fiber has 4\% loss. \subref{fig:sum_data} Input power is split between cavity eigenpolarizations. \subref{fig:error_data} The output polarization controller was set to optimize the error signal. Discrepancies between theory and data are likely due to low-passing in the photodiode.
     }%
  \label{fig:sig_exp}
\end{figure}

To optimize the error signal, the input polarization must be an equal superposition of the cavity eigenpolarizations. This is achieved by scanning the laser frequency and adjusting the input FPC until the resonant peaks in the total reflection signal---obtained from the sum of the PD signals---are balanced, as shown in figure \ref{fig:sum_data}.

The last step is to adjust the output polarization control paddles to produce an error signal matching figure~\ref{fig:error_data}.  This is done empirically; no calculation was required to produce the error signal.  Looking at the difference signal of the photodiodes, the polarization controllers are adjusted to make the peaks in the error signal the same amplitude for both resonances, and to make the signal symmetric about the frequencies which are halfway between resonances.

A typical reflected power and error signal for the ${20}\text{-m}$ fiber ring setup is shown in figure \ref{fig:sig_exp}.  Note the agreement with the theoretical model presented in section \ref{sec:err_sig}. Once the losses and coupler properties are measured, the only free parameter fit in the theoretical model is $\Delta\theta$.  It is important to note that we do not need to know what the eigenpolarizations of the resonator are, or which particular input polarization (satisfying equation \eqref{eq:in}) is being used to produce this error signal. 

\section{Conclusions}

We have presented a simple yet general method for producing an error signal for a birefringent resonator.  This method relies on the phase difference acquired by the different eigenpolarizations of the resonator on reflection, where light in one eigenpolarization is used as a phase reference for light in the other.  The method does not require any particular arrangement of eigenpolarizations, or even that they be known to the experimenter.  Instead, all that is required is the ability to tune both the polarization of the light incident on the resonator and the polarization of the reflected light before splitting it on a PBS.  Since the generation of this error signal does not require any modulation of the light or of the cavity, the electronics used to produce the error signal are remarkably simple. Further, because there is no demodulation, a low-pass filter is not needed to extract the error signal, resulting in a high-bandwidth error signal, a requirement for fast feedback.  We have demonstrated this method experimentally by producing an error signal with a ${20}\text{-m}$ long fiber ring resonator. While this method can be applied to nearly any birefringent optical cavity, this technique is especially useful for fiber-based sensors.  This method can be used to produce compact, pre-stabilized lasers, with particular relevance for fiber-based laser systems.

\bibliographystyle{osajnl}
\bibliography{FRR_Sources}




\end{document}